\documentclass[12pt]{iopart}
\usepackage{iopams}  
\usepackage{graphicx}
\usepackage{hyperref}

\begin{document}

\title[Physical properties of SrSn$_4$]{Physical properties of SrSn$_4$ single crystals}

\author{Xiao Lin,$^1$  Sergey L. Bud'ko,$^{1,2}$ German D. Samolyuk,$^3$ Milton S. Torikachvili,$^{1,4}$  and Paul C. Canfield$^{1,2}$}

\address{$^1$ Department of Physics and Astronomy, Iowa State University, Ames. IA 50011-3160, USA}
\address{$^2$Ames Laboratory, US DOE, Iowa State University, Ames. IA 50011-3020, USA}
\address{$^3$Materials Science \& Technology Division, Oak Ridge National Laboratory, Oak Ridge, TN 37831-6138, USA}
\address{$^4$Department of Physics, San Diego State University, San Diego, CA 92182-1233 USA}

\begin{abstract}
We present detailed thermodynamic and transport measurements on single crystals of the recently discovered binary intermetallic superconductor, SrSn$_4$. We find this material to be a slightly anisotropic three-dimensional, strongly-coupled, possibly multi-band, superconductor. Hydrostatic pressure causes a decrease in the superconducting transition temperature at the rate of $\approx -0.068$ K/kbar. Band structure calculations are consistent with experimental data on Sommerfeld coefficient and upper superconducting critical field anisotropy and suggest complex, multi-sheet Fermi surface formed by four bands.
\end{abstract}

\pacs{74.25.-q, 74.25.Op, 74.62.Fj, 74.70.Ad}
\submitto{\JPCM}
\maketitle

\section{Introduction}

Recent discoveries of non-oxide superconductors \cite{nag94a,cav94a,nag01a,rot08a} with respectably high transition temperatures, $T_c$, and pairing mechanisms more complex than an isotropic, single-band, phonon-mediated one, have revitalised interest in intermetallic superconductors and promoted detailed studies of relatively simple binary and ternary materials, in particularly those containing alkaline earths or light elements, that did not get enough attention in the past due to complex synthesis, air sensitivity, or were just overlooked. 

The existence of binary SrSn$_4$ phase was suggested about 30 years ago \cite{mar81a} but only recently \cite{hof03a} the crystal structure was solved and superconductivity ($T_c = 4.8$  K) was discovered. Thus SrSn$_4$ is the most recent addition to the alkaline earth - stannides group of superconductors (for SrSn$_3$, BaSn$_3$ and BaSn$_5$ the superconducting transition temperatures are $\approx 5.4$ K \cite{fas00a},  $\approx 2.4$ K \cite{fas97a}, and  $\approx 4.4$ K \cite{fas01a}, respectively). SrSn$_4$ crystallizes in the orthorhombic  structure ($Cmcm$ space group) with a unique Sr site and three different Sn sites. Of the physical properties, only low temperature and low field magnetization measured on a polycrystalline sample were reported so far \cite{hof03a}. 

In this work we report growth of single crystals of SrSn$_4$ and measurements of their thermodynamic and transport properties. Both the superconducting and normal states were characterized. The relevant results are compared with the band structure calculations.

\section{Methods}

Single crystals of SrSn$_4$  were produced by traditional solution-growth methods.\cite{can92a,can10a}  Elemental Sr and Sn were placed in a 2 ml alumina crucible (with the ratio of 3.5: 96.5). A second, catch crucible, containing silica wool was placed on the top of the growth crucible and sealed in a silica ampoule under approximately 1/3 atmosphere of argon gas. It should be noted that the packing and assembly of the growth ampoule was performed in a glovebox with nitrogen atmosphere. The sealed ampoule was heated up to 800$^\circ$ C over several hours, then cooled to 370$^\circ$ C , and slowly cooled to 260$^\circ$ C over 36 h. At this temperature, the crucible was removed from the furnace and the remaining liquid was decanted. The crystals grow as elongated blades or rods of a few mm length and sub-mm the two other dimensions. The crystals are air-sensitive, and have to be kept in the nitrogen glovebox.  An effort was made to minimize the exposure to air in the process of handling and shaping the samples for the  thermodynamic and transport measurements.

Powder x-ray diffraction data on both non-oxidized and oxidized sample were collected by a Rigaku Miniflex diffractometer with Cu K$\alpha$ radiation at room temperature. The diffraction pattern of the non-oxidized SrSn$_4$ was taken from the powder of SrSn$_4$ single crystals that were ground in a glovebox. The sample powder was sealed by Kapton film during the measurement to protect it from oxidization. The same powder was used for the oxidized x-ray diffraction after a seven-hour exposure to the air. 

Temperature- and magnetic field - dependent dc magnetization was measured using a Quantum Design MPMS-5 SQUID magnetometer. Electrical resistance was measured in a Quantum Design PPMS instrument using a conventional four-probe geometry. Platinum wires were attached to the sample using EpoTek H20E silver epoxy with the current flowing approximately along the longest dimension of the crystal ($a$-axis). The epoxy was cured in air at 120$^\circ$ C for $\sim 20$ min. This procedure resulted in some surface oxidation, but seems to preserve the structural integrity and electrical current continuity for the resistance samples.  Heat capacity was measured in a PPMS instrument using a relaxation technique.  Low field dc magnetization under pressure up to $\sim 10$ kbar was measured in MPMS-5 using a commercial, HMD, Be-Cu piston-cylinder pressure cell \cite{hmd}  with Daphne oil 7373 as a pressure medium and superconducting Pb as a low temperature pressure gauge \cite{eil81a}. 

\section{Results and Discussion}
\begin{figure}[tbp]
\begin{center}
\includegraphics[angle=0,width=120mm]{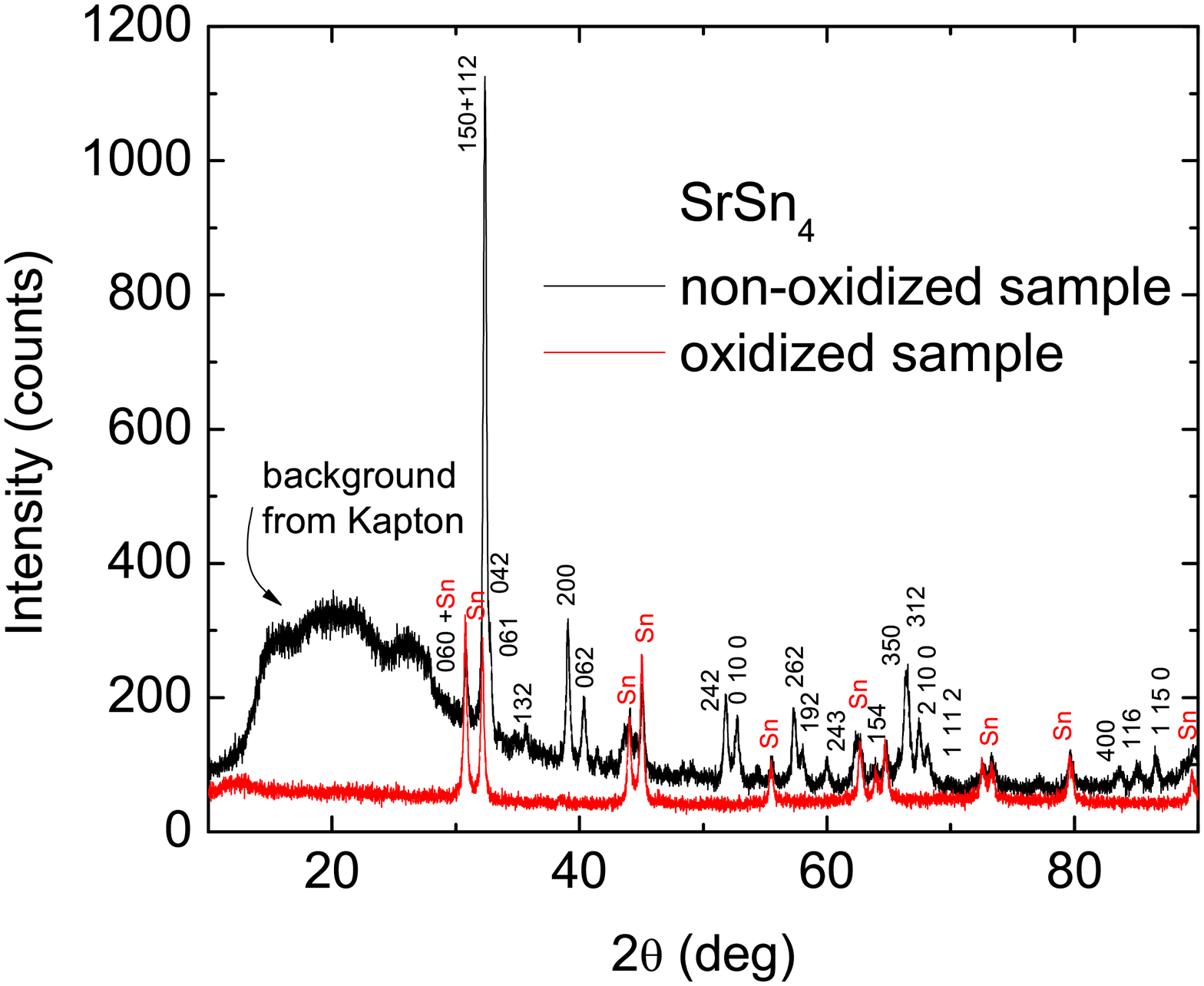}
\end{center}
\caption{(Color online) Comparison of the x-ray patterns taken on non-oxidized and oxidized powdered SrSn$_4$ single crystals. Peaks that belong to the SrSn$_4$ are labeled with their $h~k~l$ values. Note: the only difference between the two runs were (i) removal of Kapton film and (ii) 7 hours exposure to air. Note: second phase Sn peaks (from small amounts of residual Sn-flux) are essentially unchanged by exposure to air.} \label{F1}
\end{figure}

Powder x-ray diffraction (Fig. \ref{F1}) on the non-oxidized sample confirms that the synthesized crystals are SrSn$_4$ with $Cmcm$ structure. The obtained lattice parameters are $a$ = 4.625(2) \AA,~ $b$ = 17.384(6) \AA~and $c$ = 7.077(3) \AA, consistent with the literature data \cite{hof03a}. A few lines from residual Sn flux are also visible. In contrast, the diffraction peaks of SrSn$_4$ all disappear in the oxidized sample's diffraction pattern, leaving only Sn peaks, with their intensities remaining basically unchanged (Fig. \ref{F1}). These results imply that the oxidation of the powdered SrSn$_4$ results in the phase that is too small or disordered to diffract.
 
\begin{figure}[tbp]
\begin{center}
\includegraphics[angle=0,width=120mm]{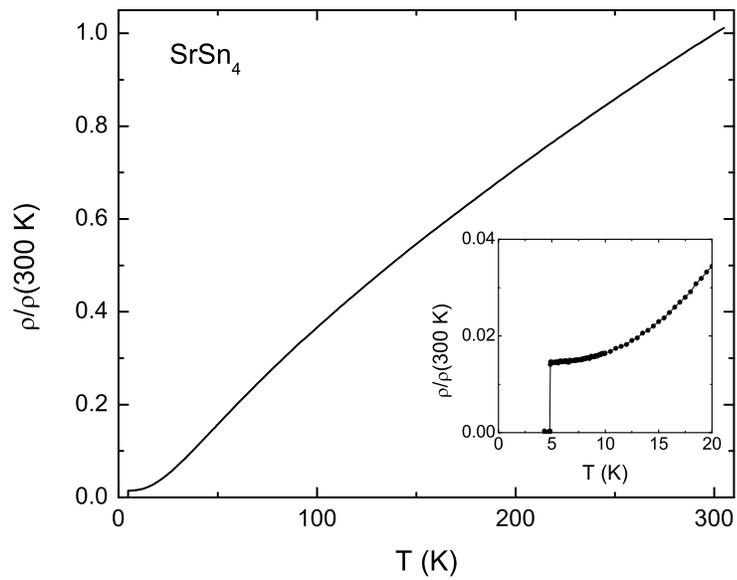}
\end{center}
\caption{The temperature-dependent, normalized resistivity of SrSn$_4$. Inset: low temperature data showing the superconducting transition.} \label{F2}
\end{figure}

Temperature dependent resistivity data for SrSn$_4$ are presented in Fig. \ref{F2}. Since the sample was of somewhat irregular shape, the data are presented normalized to the room temperature value. The room temperature resistivity, to within factor of two, is approximately 200 $\mu\Omega$ cm. The resistivity data show metallic ($d\rho/dT > 0$) behavior with rather high residual resistivity ratio, $RRR = \rho(300$ K$)/\rho(5$ K$) \sim 70$. The superconducting transition, with an onset at $\sim 4.9$ K, is clearly seen in the data. 

\begin{figure}[tbp]
\begin{center}
\includegraphics[angle=0,width=120mm]{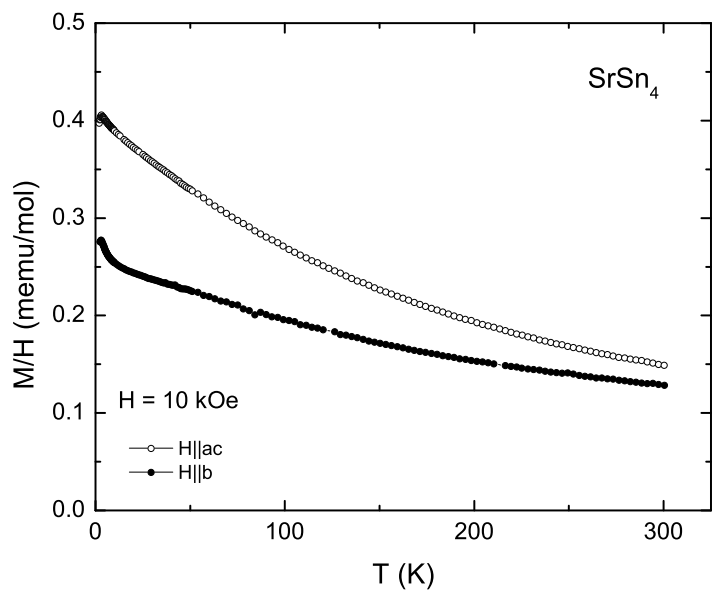}
\end{center}
\caption{ Anisotropic temperature dependent susceptibility, $M/H$, of SrSn$_4$.} \label{F3}
\end{figure}

Anisotropic temperature-dependent dc susceptibility, $M/H$,  is shown in Fig. \ref{F3}.  Normal state, high field susceptibility, is paramagnetic and anisotropic, with $(M/H)_{ac} > (M/H)_b$ in the whole temperature range. Anisotropic $M(H)$ measured at 5 K is close to linear,  possibly with small para- or ferromagnetic contribution that saturates by $\sim 5$ kOe at $\sim 2~10^{-4}~\mu_B$/f.u. value.   The temperature dependence of the susceptibility is somewhat surprising. The high temperature part of the anisotropic $M/H(T)$ in principle can be fitted with a modified Curie - Weiss law,  $\chi = \chi_0 + C/(T-\Theta)$, (resulting in rather large, $0.7-0.9 \mu_B$, effective moment and large, negative, order of -200 K, values of $\Theta$).  At this point we have doubts that such local-moment-like picture has any merit.  More work is required to determine if this temperature dependence is intrinsic and what is the mechanism behind it. 

\begin{figure}[tbp]
\begin{center}
\includegraphics[angle=0,width=120mm]{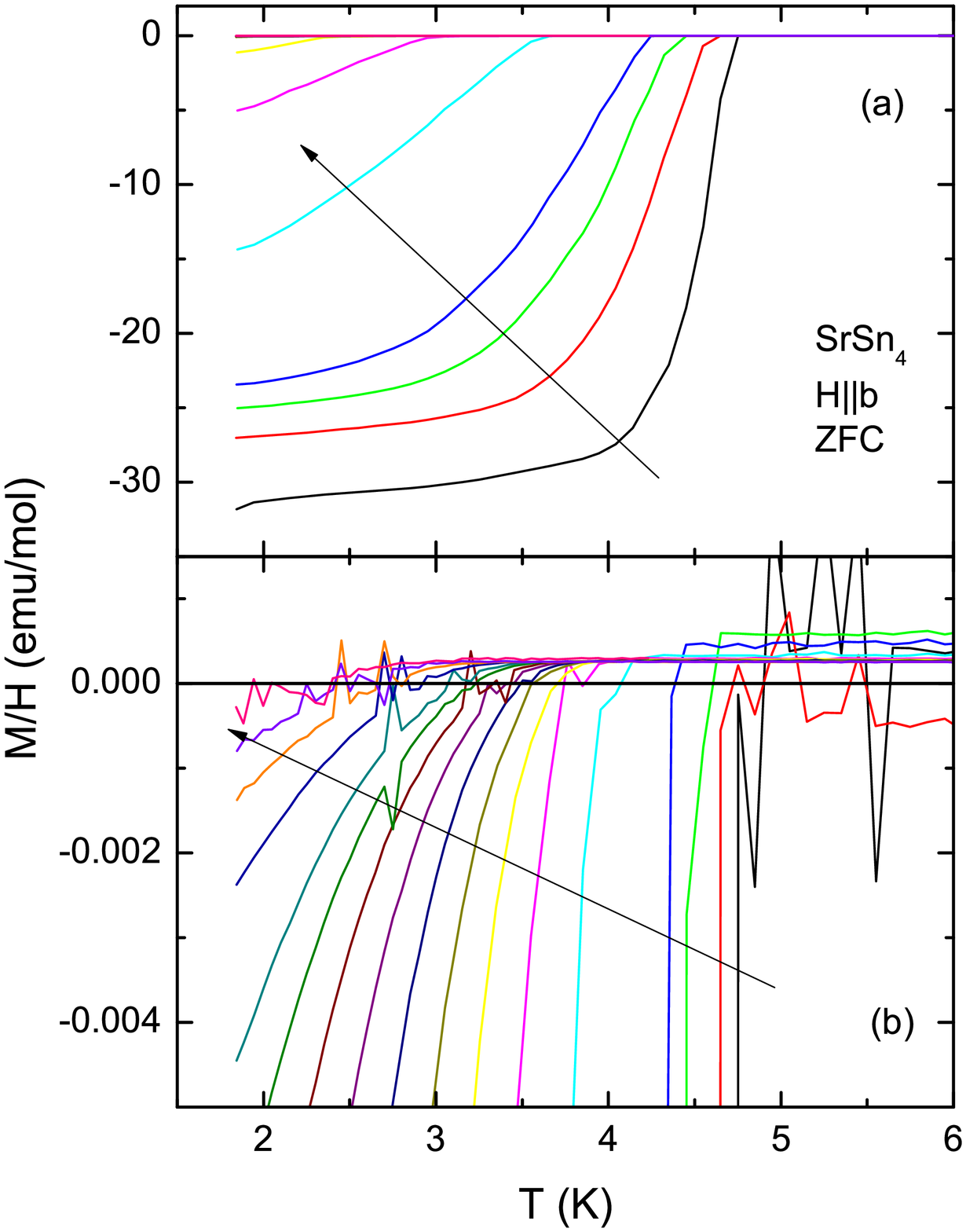}
\end{center}
\caption{(Color online) (a) ZFC temperature-dependent susceptibility ($H \| b$) of SrSn$_4$ measured at 25, 50, 75, 100, 200, 300, 400, 500, 600, 700, 800, 900, 1000, 1200, 1400, 1600, and 1800 Oe (arrow shows the direction of increasing applied field); (b) data of panel (a) enlarged near $M/H = 0$.} \label{F4}
\end{figure}

\begin{figure}[tbp]
\begin{center}
\includegraphics[angle=0,width=120mm]{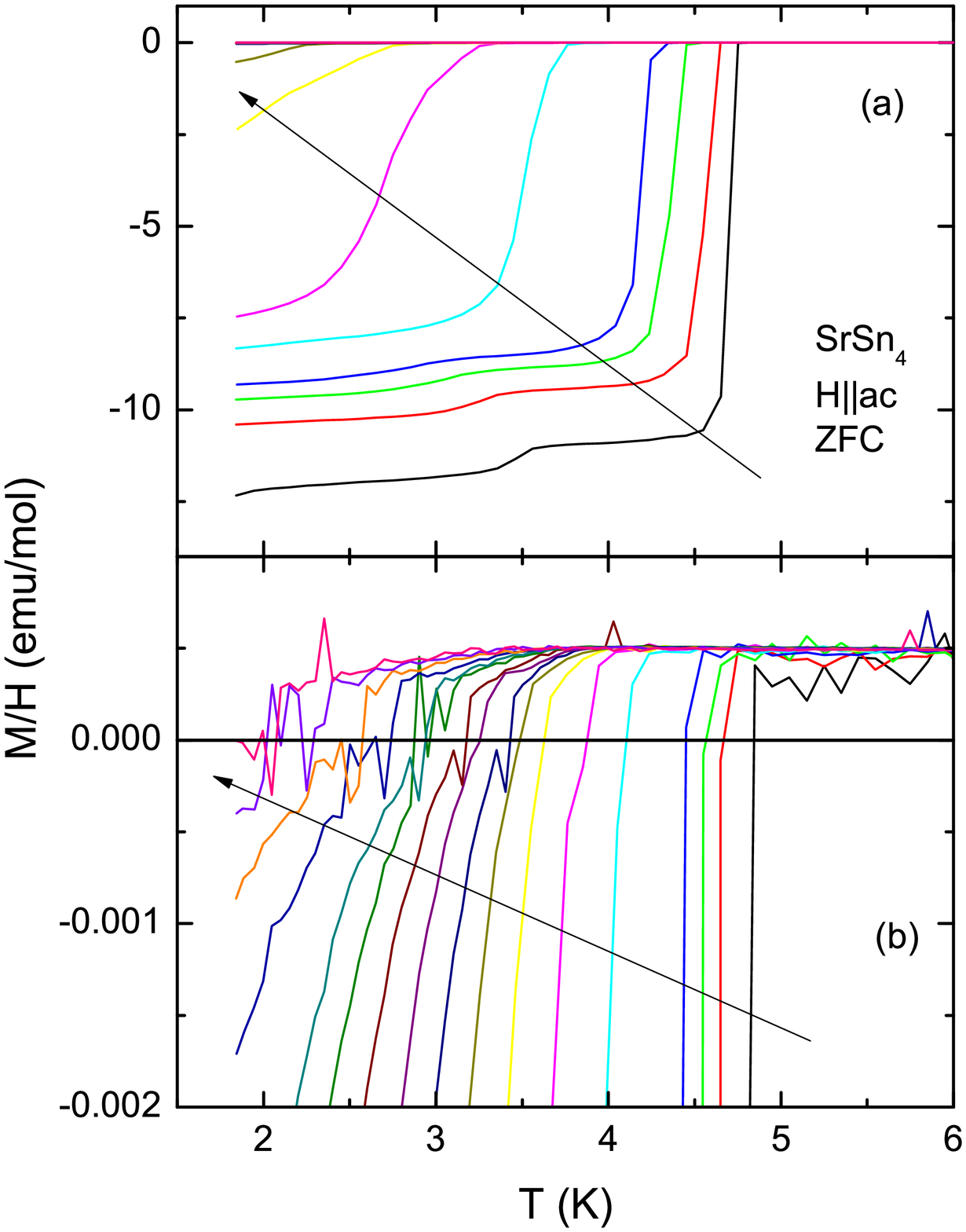}
\end{center}
\caption{(Color online) (a) ZFC temperature-dependent susceptibility ($H \| ac$) of SrSn$_4$ measured at 25, 50, 75, 100, 200, 300, 400, 500, 600, 700, 800, 900, 1000, 1200, 1400, 1600, and 1800 Oe (arrow shows the direction of increasing applied field); (b) data of panel (a) enlarged near $M/H = 0$.} \label{F5}
\end{figure}

\begin{figure}[tbp]
\begin{center}
\includegraphics[angle=0,width=90mm]{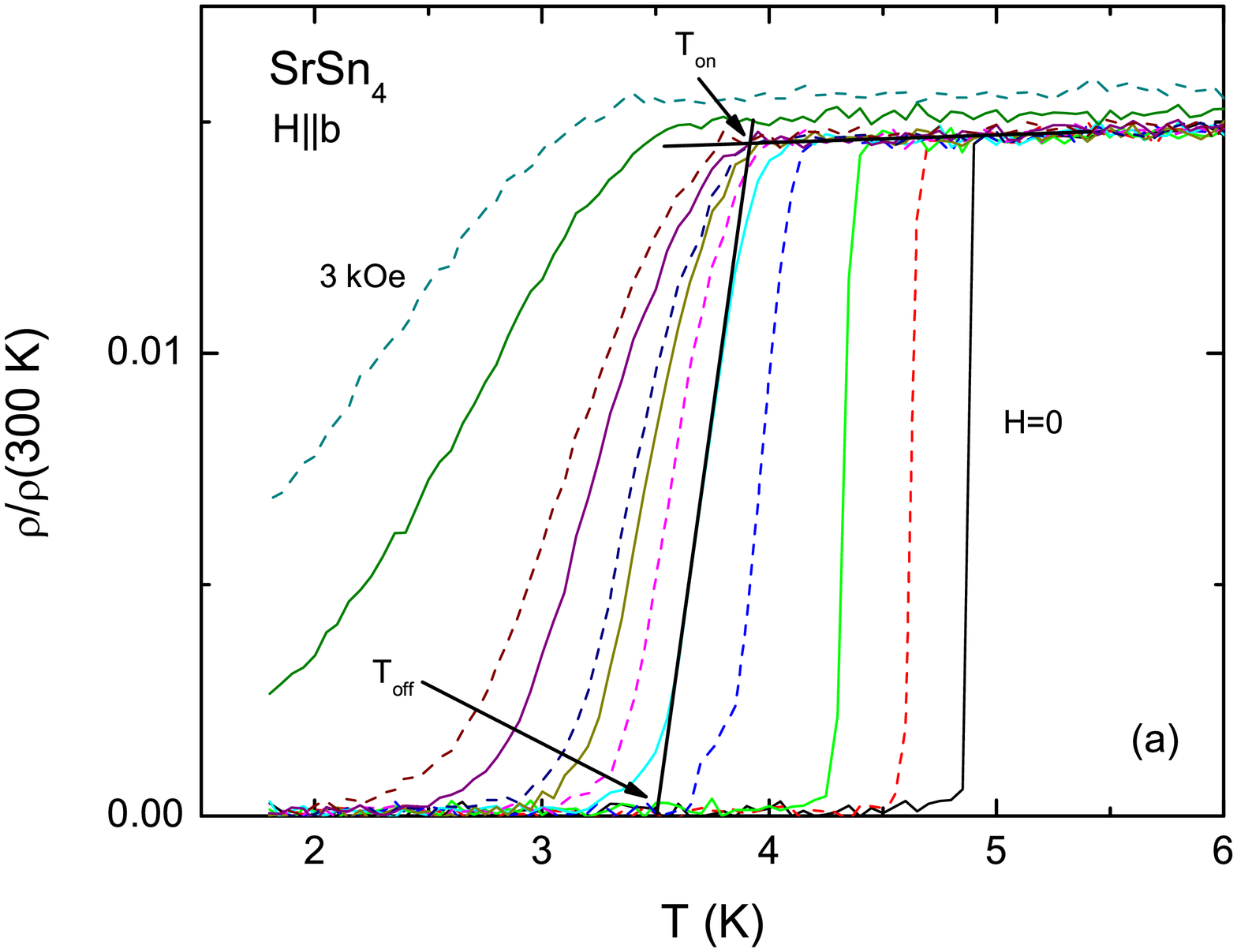}
\includegraphics[angle=0,width=90mm]{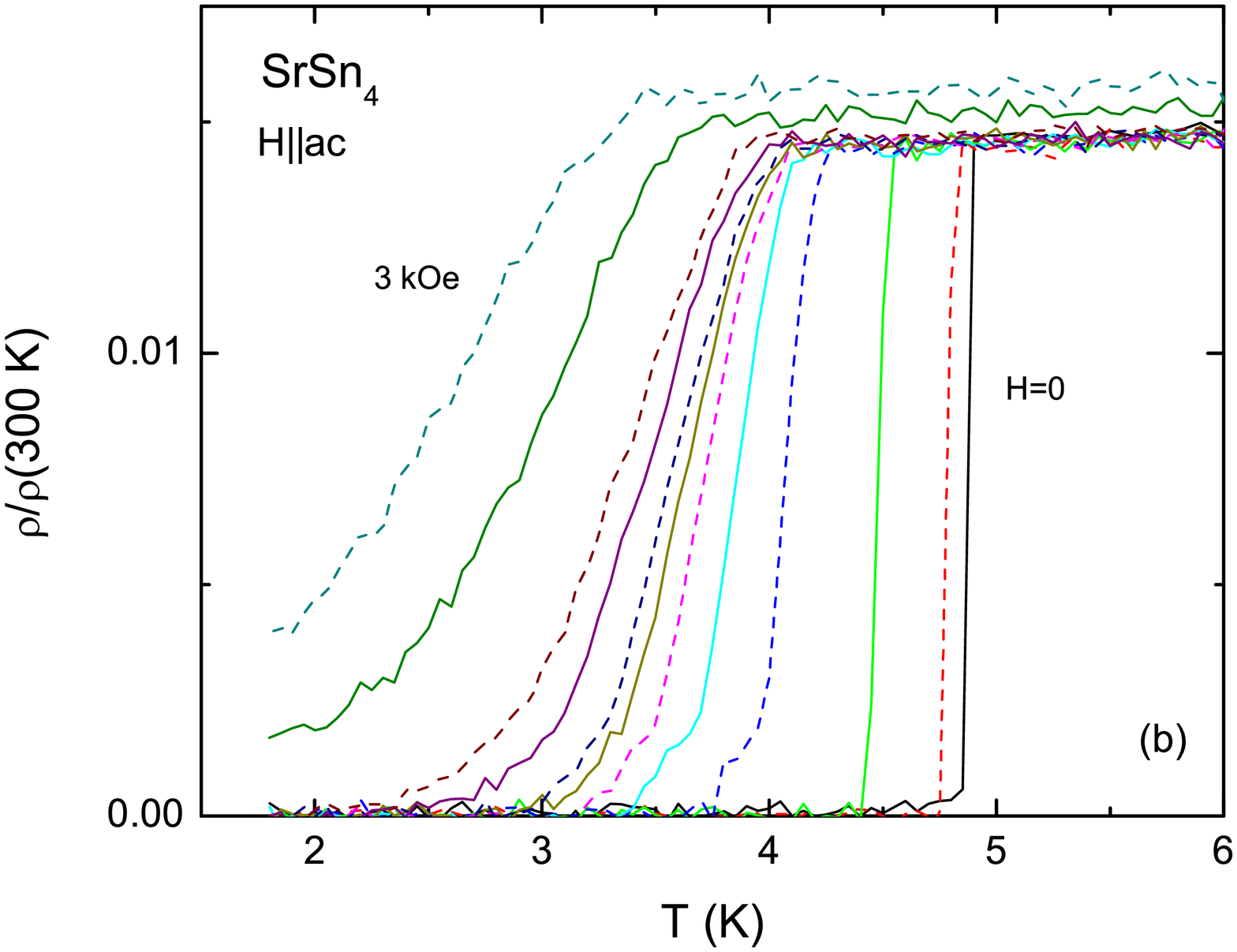}
\end{center}
\caption{(Color online ) Low temperature resistivity  of SrSn$_4$ measured at 0, 50, 100, 200, 300, 400, 500, 600, 800,  1000, 2000,  and 3000 Oe. (a) $H \| b$, (b) $H \| ac$. Criteria for $T_{onset}$ and $T_{offset}$ are shown for the $H = 300$ Oe, $H \| b$ data.} \label{F45a}
\end{figure}

\begin{figure}[tbp]
\begin{center}
\includegraphics[angle=0,width=120mm]{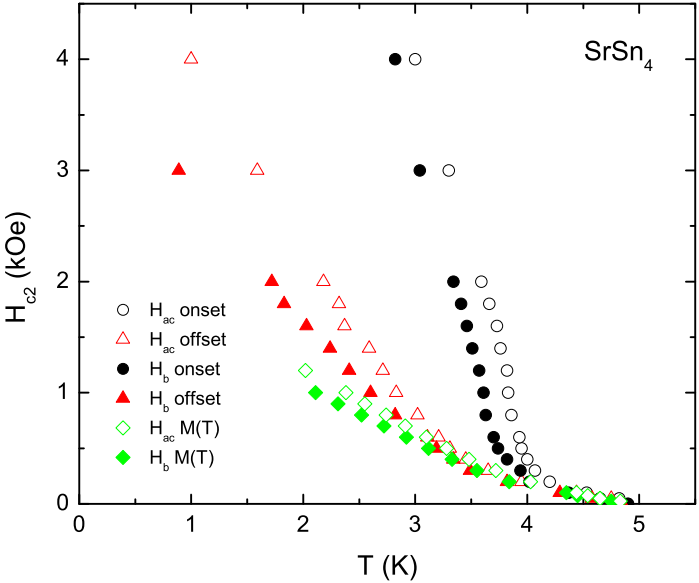}
\end{center}
\caption{(Color online) Anisotropic upper critical field of SrSn$_4$ from magnetization and magnetotransport measurements.} \label{F6}
\end{figure}

Superconductivity with the onset of $\sim 4.8$ K  is clearly seen in the  low field zero-field-cooled (ZFC) susceptibility data,   Figs. \ref{F4} and \ref{F5},  (no corrections for demagnetization factor were employed). To infer an anisotropic upper superconducting critical field for SrSn$_4$, a set of low field ZFC susceptibility measurements were performed (see Figs. \ref{F4},\ref{F5}). A criterion of $M=0$  was chosen for the upper critical field. Alternatively, anisotropic $H_{c2}(T)$ can be evaluated from the shifts of resistively measured superconducting transitions in different applied magnetic fields (Fig. \ref{F45a}). Onset and offset criteria are used for the resistivity data (Fig. \ref{F45a}(a)).  The resulting anisotropic $H_{c2}(T)$ curves are shown in Fig. \ref{F6}. As it often the case, $T_{offset}$ criterion agrees quite well with $T_{onset}$ criterion for magnetization. Albeit different measurements and criteria give somewhat different $H_{c2}(T)$ curves, several features are observed in all of them.  A distinct upward curvature is clearly seen for both orientations (this upturn is particularly sharp in onset resistivity data).  This feature might be a signature of multiband superconductivity \cite{shu98a}. The upper critical field of SrSn$_4$ is rather small, clearly below 10 kOe at $T = 0$. The anisotropy of $H_{c2}$ is rather small, it does not exceed $\sim 1.5$.  $H_{c2}(T)$ curves from magnetization and offset resistive data are similar. Taken together these data suggest that this material might have complex superconductivity and/or superconducting $H - T$ phase diagram. It has to be noticed that $T_c$ and the superconducting critical fields found for SrSn$_4$ in this work are significantly different from that for elemental Sn used as flux  ($T_c(Sn) \approx 3.7$ K, and $H_c(Sn,  T= 0) = 305$ Oe), which rules out traces of Sn flux in the crystals as the source of the superconducting behavior \cite{fin65a}.

\begin{figure}[tbp]
\begin{center}
\includegraphics[angle=0,width=120mm]{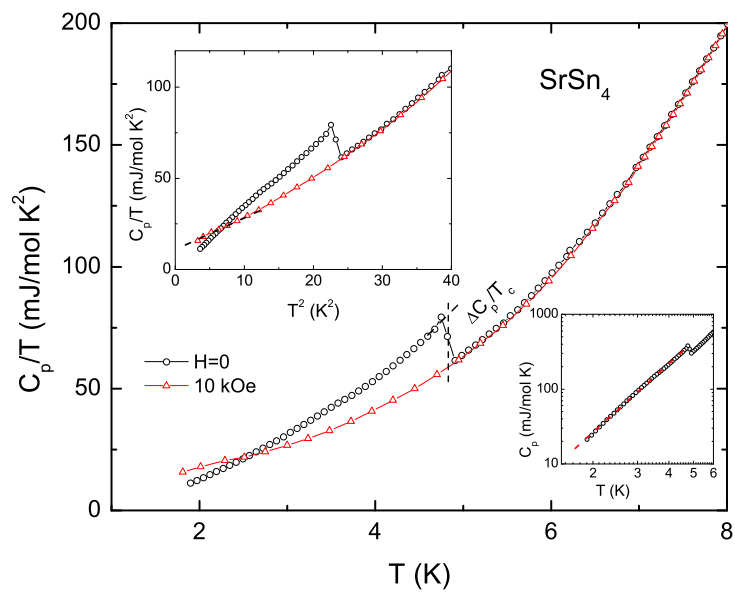}
\end{center}
\caption{(Color online) Low temperature heat capacity of SrSn$_4$ plotted as $C_p$ vs $T$ in zero and 10 kOe ($H \| b$) applied field. Upper left inset: the same data plotted as $C_p$ vs $T^2$, dashed line - extrapolation of the low temperature linear region of the 10 kOe data. Lower right inset: $H = 0$ $C_p(T)$ data plotted on {\it log - log} scale, dashed line corresponds to $C_p \propto T^3$.} \label{F7}
\end{figure}

The low temperature heat capacity of SrSn$_4$ is shown in Fig. \ref{F7}. A 10 kOe applied field is enough to completely suppress superconductivity (Fig. \ref{F6}) without changing the normal state properties. From the low temperature, in-field, heat capacity data,  the Sommerfeld coefficient for SrSn$_4$ is estimated to be $\gamma \approx 9.7$ mJ/mol K$^2$, and the Debye temperature  $\Theta \approx 170$ K. The superconducting transition is clearly seen in zero-field heat capacity data. The jump at $T_c$ is $\Delta C_p/T_c \approx 20.8$ mJ/mol K$^2$ that gives corresponds to $ \Delta C_p/\gamma T_c \approx 2.1$  This value is higher than the canonical  1.43 value expected for isotropic weakly coupled BCS superconductor and suggests that SrSn$_4$  is a strongly coupled superconductor \cite{car90a}. Finally, the $C_p(T)$ behavior in the superconducting state (Fig. \ref{F7}, lower right inset) appears to be non-exponential and reasonably well described by  $C_p \propto T^3$ function. If intrinsic, such dependence might point to deviations from isotropic single band superconductivity for this material. 

\begin{figure}[tbp]
\begin{center}
\includegraphics[angle=0,width=120mm]{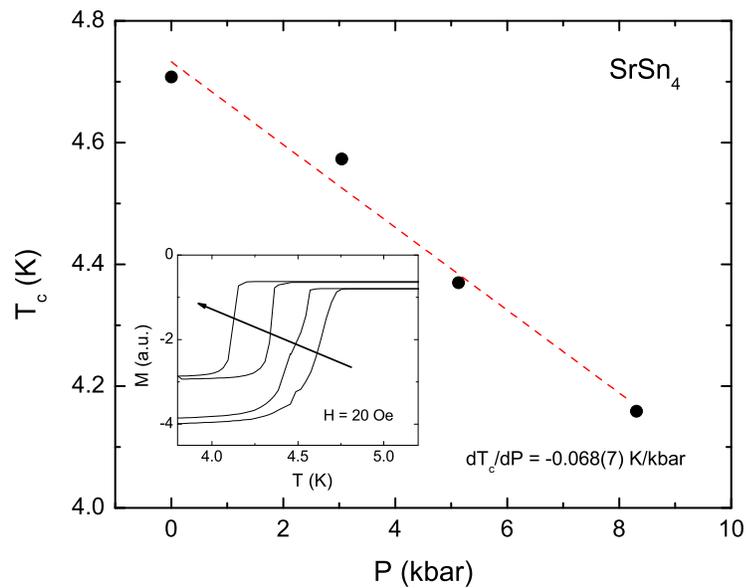}
\end{center}
\caption{(Color online) Pressure dependence of the superconducting transition temperature of SrSn$_4$. Dashed line - linear fit. Inset - low field magnetization under pressure, arrow points in the direction of increasing pressure.} \label{F8}
\end{figure}

The superconducting transition temperature of SrSn$_4$  linearly decreases under pressure up to $\sim 8$ kbar (Fig. \ref{F8}). The pressure derivative $dT_c/dP = - 0.068 \pm 0.007$ K/kbar, is rather small, similar in sign and order of magnitude to those measured for a number of  elemental and binary superconductors \cite{bra65a}. Such pressure dependence is possibly the result of rather weak dependence of the density of states on energy near the Fermi level as well as possibly opposing changes to $T_c$ caused by shift in the phonon spectrum by hydrostatic pressure.\\

\begin{figure}[tbp]
\begin{center}
\includegraphics[angle=270,width=120mm]{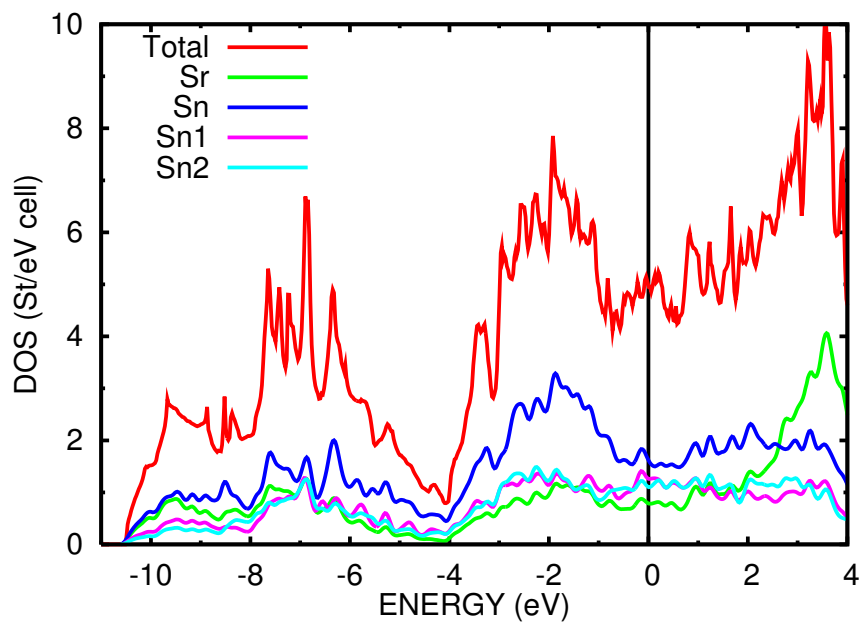}
\end{center}
\caption{(Color online) The calculated total density of states together with partial density of states of symmetry non-equivalent atoms. The zero energy corresponds to Fermi level ($E_F$)} \label{F9}
\end{figure}

The electronic structure calculation has been done within generalized gradient approximations of density functional theory  using the {\sc quantum espresso} package \cite{gia09a}.We have employed plane-wave basis set and ultrasoft pseudo-potential \cite{van90a} and the Perdew, Wang \cite{per92a} exchange-correlation functional. The Brillouin zone (BZ) summations were carried out over a 16x16x8 BZ grid with Gaussian boarding 0.02 Ry in self-consistent calculation and summation over tetrahedrons for electronic density of states calculation. The plane wave energy cut off at 30 Ry allows an accuracy of 0.2 mRy/atom. The calculated density of states (DOS) is presented in Fig. \ref{F9}. Similar to the results published in \cite{hof03a,wen11a}, the Fermi level ($E_F$) is located near the edge of a local maximum and close to local minimum. However the local minimum is less pronounced compared to LMTO results \cite{hof03a,wen11a}. The value of DOS at $E_F$ ($\approx 5$ St/eV cell) is pretty high and the corresponding $\gamma_{band} \approx 5.8$ mJ/mol K$^2$. Comparison with the experimental value, $\gamma_{exp} = (1 + \lambda) \gamma_{band}$,  yields the value of the enhancement factor (due to electron - electron and electron - phonon interactions) $\lambda \approx 0.7$, that appears to be comparable to the enhancement observed in alkali metals due to electron - phonon interactions \cite{gri81a}   The almost equal distribution of total DOS between all atoms in the compound points to strong hybridization of valence electrons (that might contribute to the value of $\lambda$). The detailed analysis of character of electrons bonds have been provided in \cite{hof03a}.

\begin{figure}[tbp]
\begin{center}
\includegraphics[angle=270,width=120mm]{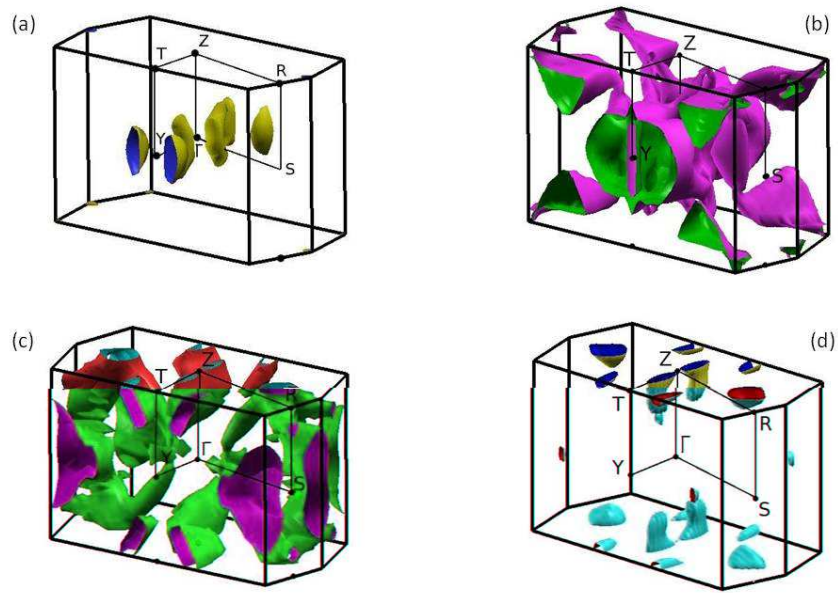}
\end{center}
\caption{(Color online) Theoretical Fermi surface of SrSn$_4$. The sheets are formed by bands 4-7 [panels (a) to (d), respectively]. Xcrysden visualization program was used to plot the FS \cite{kok99a}.} \label{F10}
\end{figure}

According to the calculations, a Fermi surface (FS) of SrSn$_4$ is formed by multiple sheets formed by four bands (Fig. \ref{F10}). The FS topology is complex: bands 4 and 7 give contribution in form of a number of rather small, distorted, ellipsoids, whereas bands 5 and 6 contribute to the FS in the form of multiply-connected "monsters". There appears to be no dominating low-dimensional, cylindrical, FS parts. The small $H_{c2}$ anisotropy is consistent with this three-dimensional FS. Similarly, no profound anisotropy in the electrical transport is expected.
 
\section{Summary}
Detailed thermodynamic and transport measurements were performed on single crystals of SrSn$_4$ with $RRR \sim 70$. The normal state magnetic susceptibility is paramagnetic and slightly anisotropic. SrSn$_4$ can be characterized as strongly-coupled three dimensional superconductor with $T_c \approx 4.8$ K, and the upper critical field of the order of 2 kOe. The anisotropy of $H_{c2}$ does not exceed 1.5 and is slightly temperature-dependent. $T_c$ decreases slowly under hydrostatic pressure up to $\sim10$ kbar. The heat capacity and $H_{c2}(T)$ data suggest that superconductivity in this material may be more complex than isotropic BCS. Band structure calculations of the DOS at $E_F$ are consistent with experimental observations, suggesting the strong hybridization of valence electrons and yielding three-dimensional Fermi surface with multiple sheets originating from four bands.

Since SrSn$_4$ crystals have high $RRR$ and easily accessible $T_c$, detailed studies of the symmetry of the superconducting state by low temperature thermal transport and/or penetration depth measurements could be of interest.

\ack
This work was carried out at the Iowa State University and supported by the AFOSR-MURI grant \#FA9550-09-1-0603 (X. Lin, M. S. Torikachvili and P. C. Canfield). Part of this work was performed at Ames Laboratory, US DOE, under contract \# DE-AC02-07CH 11358 (S. L. Bud'ko). S. L. Bud'ko also acknowledges partial support from the State of Iowa through Iowa State University and valuable input from Willy and Tom Tuttle. Research at Oak Ridge National Laboratory  (G. D. Samolyuk) was sponsored by the Division of Materials Sciences and Engineering, Office of Basic Energy Sciences, US DOE.  M. S. Torikachvili was supported in part by the National Science Foundation under DMR-0805335.

\clearpage

\clearpage

\clearpage

\clearpage

\clearpage

\clearpage

\clearpage

\clearpage

\clearpage

\clearpage

\clearpage

\end{document}